\documentclass[prd, twocolumn, nofootinbib]{revtex4}

\usepackage{epsfig} 
\usepackage{amsmath}

\newcommand{\beq}{\begin{equation}}
\newcommand{\eeq}{\end{equation}}
\newcommand{\beqa}{\begin{eqnarray}}
\newcommand{\eeqa}{\end{eqnarray}}

\newcommand{\ls}{\mathrel{\raise0.27ex\hbox{$<$}\kern-0.70em \lower0.71ex\hbox{{
$\scriptstyle \sim$}}}}

\begin{document} 

\title{Being PC: Principal Components and Dark Energy} 
\author{Roland de Putter and Eric V.\ Linder} 
\affiliation{Berkeley Lab \& University 
of California, Berkeley, CA 94720, USA} 
\date{\today}

\begin{abstract} 
Principal component analysis is considered as an addition to the well-tested 
parametrization $w(a)=w_0+w_a(1-a)$ for the dark energy equation 
of state.  This brief note cautions against some unjustified 
assumptions in interpretation of PCA calculations, giving quantified 
examples. 
\end{abstract} 

\maketitle

\section{Introduction \label{sec:intro}}

Dark energy is a premier mystery of cosmology and high energy physics. 
To address this, NASA and the Department of Energy (DOE) are funding the 
Joint Dark Energy Mission (JDEM).  Because the physical nature of 
dark energy is so unknown, it is challenging to quantify simply and 
accurately the science requirements for learning about the physical 
origin.  To assess approaches to a figure of merit for the dark 
energy science reach of different JDEM architectures, NASA and DOE 
formed the Figure of Merit Science Working Group (FOMSWG). 

FOMSWG found in a preliminary report \cite{fomswg} that 
the figure of merit used by the Dark Energy Task Force \cite{detf}, 
given by the inverse area of the likelihood contour in the dark energy 
equation of state plane $w_0$-$w_a$ was reasonable for the task.  Here the 
equation of state (EOS) as a function of scale factor is given by 
$w(a)=w_0+w_a(1-a)$.  
This form has been tested for physical accuracy and against bias, and has 
been shown to faithfully reproduce relations for observables such as 
distances and Hubble parameters to an accuracy of $10^{-3}$ \cite{calib}, 
better than needed for JDEM. 

These results, and especially the calibration relations of \cite{calib}, 
should substantially allay concerns about using a particular functional 
form.  Recall that the current issue is how to {\it project\/} simulated 
constraints of various JDEM scenarios; once the data are in hand one will 
carry out the analysis through a diversity of methods, and test 
specific models directly against the data without an intermediate form. 
Nevertheless, one can reasonably consider an {\it additional\/} 
(not replacement) approach to build confidence that the conclusions 
on science design were robust.  An example is principal component 
analysis (see, e.g., \cite{hutstar,albern,deplpca,mort} for 
application to dark 
energy), which has the capability of covering a wide variety of 
functional forms.  

The assessment of whether principal components (PCs) add new insights 
to JDEM design projections depends on appropriate scientific criteria. 
Here we present brief cautions about possible oversimplifications of 
interpretation, titling the following section headings with some of the 
possible misunderstandings.

\section{Uncertainties are Small, So We Know the Answer?} 

The principal components compose the EOS through 
\beq 
w(a)-w_{\rm b}(a)=\sum_i \alpha_i\,e_i(a), \label{eq:pc} 
\eeq 
where $w_{\rm b}$ is the baseline EOS to compare to (e.g.\ $w=-1$), 
$e_i$ are the eigenmodes 
of the Fisher matrix for the particular experiment, and $\alpha_i$ 
are the mode coefficients.  

Given the information from some 
(simulated) experiment, one forms the Fisher information matrix and 
diagonalizes it.  The rows of the diagonalization matrix are the 
eigenvectors, or modes $e_i(a)$.  If 
the Fisher matrix was formed with respect to the baseline model, then 
the expectation value of the coefficients $\langle\alpha_i\rangle=0$.  
The rms about the mean is $\sigma_i\equiv\sigma(\alpha_i)$, i.e.\ the 
inverse square root of the eigenvalues. 

One of the key misapprehensions of PCA is the physical interpretation 
of the uncertainties on the amplitudes $\alpha_i$ of each mode.  
Although PCs are often ordered by the uncertainties, these values 
$\sigma_i$ have no physical meaning by themselves.  
One has to interpret their magnitude in terms of some distance between 
models.  One possibility is adopting the range of $w\in [-1,0]$, valid 
for many models (with the upper limit to avoid disrupting early 
matter domination), and considering values of $\sigma_i$ as useful 
if they are smaller than one.  For example, FOMSWG imposes priors on 
the EOS such that $\sigma_i$ cannot exceed one.  This prescription is 
equivalent to a redefinition 
\beq 
\sigma_{\rm true} \rightarrow \sigma_{\rm FOM}= 
\frac{\sigma_{\rm true}}{\sqrt{1+\sigma_{\rm true}^2}}\,. \label{eq:sigscale}
\eeq 

One might choose to form a figure of merit given as the reciprocal of 
the product of all the $\sigma_i$'s.  This appears similar to a  
multi-dimensional extension (cf.\ \cite{albern}) of the DETF figure 
of merit written in terms of 
$1/[\sigma(w_p)\,\sigma(w_a)]$ where $w_p$ is the pivot EOS value. 
However, none of this addresses the meaning.  Since every PC by 
construction has an uncertainty smaller than or equal to one, does 
this mean that all give physical insight?  As argued by 
\cite{linhut,leach,deplpca}, the $\sigma_i$'s alone capture no physics. 
The imposition of a prior does define a goal for ``smallness'' of 
the uncertainty, if somewhat artificially.  However, just as in 
most astronomy, the true information is held by the 
signal-to-noise ratio, 
\beq 
S/N=\left[\sum \frac{\alpha_i^2}{\sigma_i^2}\right]^{1/2}, \label{eq:snr}
\eeq 
where each mode contributes $\alpha_i/\sigma_i$ worth of $S/N$ 
\cite{kadota,leach}.  This is equivalent to the $\chi^2$ comparison of 
mode coefficients between a model and the baseline.  So the important 
quantity is not $\sigma_i$ but $\sigma_i/\alpha_i$.  

Consider if one misestimated the value of some PC coefficient $\alpha_i$ 
by $\sigma_i$ (i.e.\ $1\sigma$) -- would it change the physics conclusion? 
In particular, the cosmological constant 
corresponds to $\alpha_i=0$ for all PCs.  If some $\alpha_i$ is 
within $1\sigma$ (or some other confidence level) of zero, then 
that PC is not effectively contributing to distinction of the 
physics from a cosmological constant, since there is a high statistical 
probability that $\alpha_i$ is consistent with zero.  (This can of course 
be generalized to differences between any two models.)  

Therefore, how many PCs truly contribute to distinguishing some 
model from the cosmological constant $\Lambda$?  We test this 
for a continuum of models, over various model classes, assuming 
an imaginary experiment giving 0.3\% 
distance measurements between $z=0-3$, plus Planck CMB 
information.  Every one of 36 PCs between $a=0.1-1$ 
have $\sigma_i$'s less than one, by the construction of 
Eq.~(\ref{eq:sigscale}).  
However, when we carry out the test for various model parameters, 
and model classes, below, we find generically that only $\sim$2-3 
PCs have $\sigma_i/\alpha_i<1/2$, i.e.\ a $2\sigma$ deviation from 
the cosmological 
constant, even with the extraordinary assumptions of experimental 
accuracy.  This is in agreement with the multiple studies of this 
issue by \cite{linhut}. 

\begin{figure}[!htb]
\begin{center}
\psfig{file=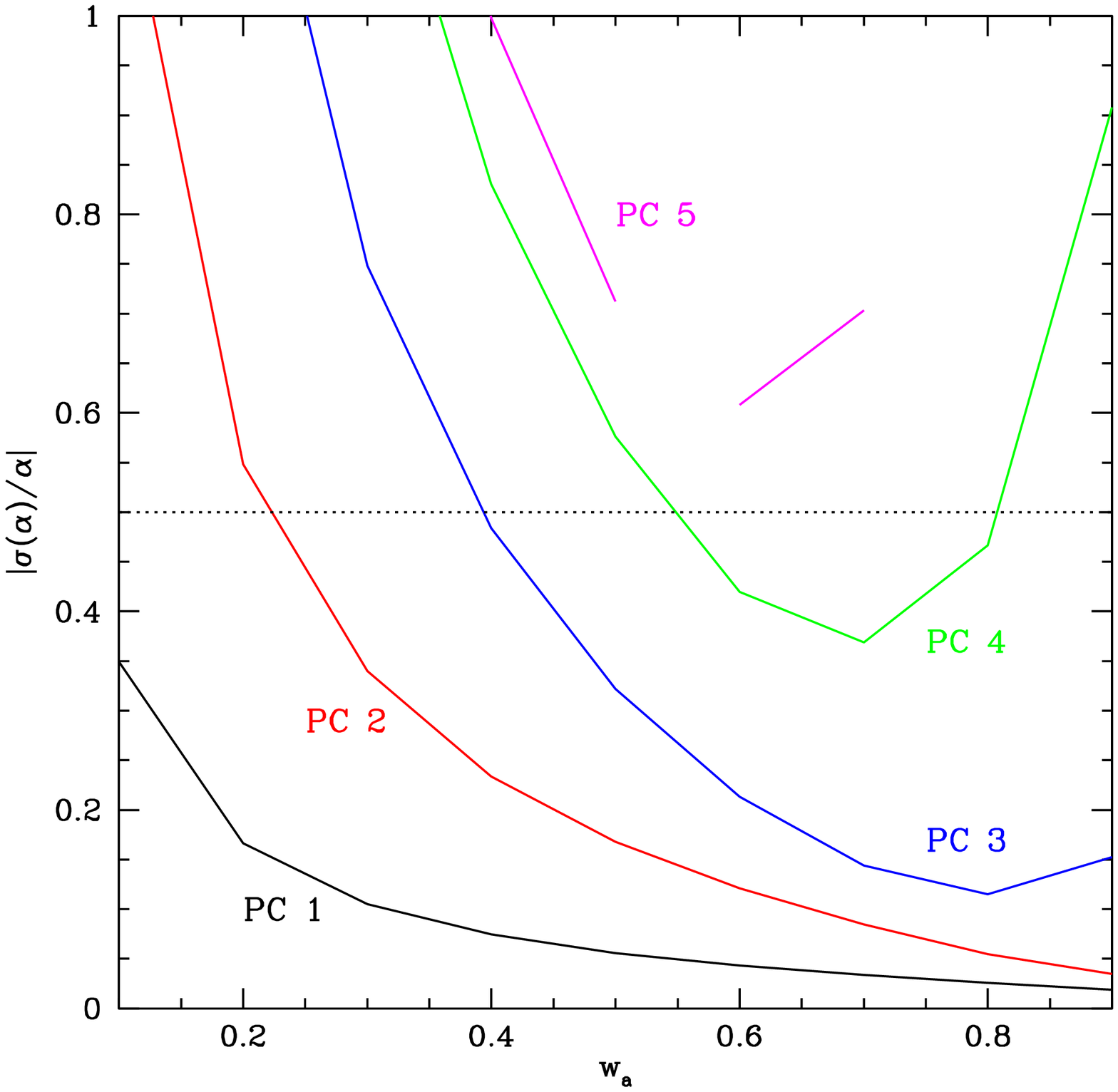,width=3.4in}
\psfig{file=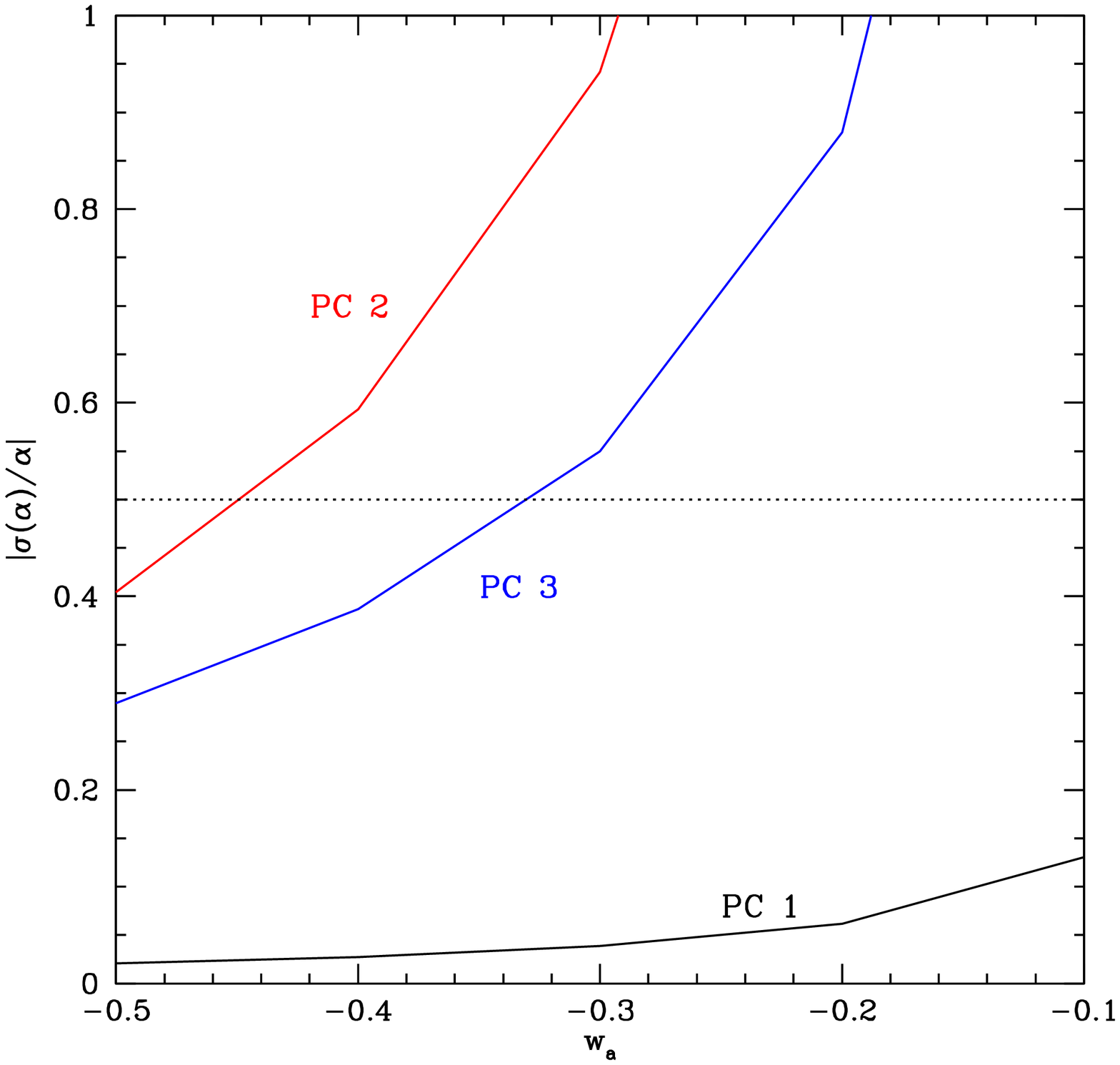,width=3.4in}
\caption{Noise-to-signal ratios $\sigma_i/\alpha_i$ vs.\ deviation
from the cosmological constant for models with $w(a)=-1+w_a(1-a)$ in
the freezing class (top), and $w(a)=(-1-w_a)+w_a(1-a)$ in the thawing 
class (bottom). 
Approaching the cosmological constant, $w_a\to0$, the models become 
difficult to distinguish from $\alpha=0$, i.e.\ $\Lambda$.  
Only 3-5 PCs have $S/N>1$, for any of the models. 
The intersection of the horizontal line 
with a PC shows where that mode gives a $2\sigma$ indication of difference 
from $\Lambda$.  For $|w_a|<0.4$, only 2 PCs meet this criterion. 
(Note breaks in the PC5 curve come from $\alpha_5$ switching sign.)  
}
\label{fig:sigalf}
\end{center}
\end{figure}

\begin{figure}[!htb]
\begin{center}
\psfig{file=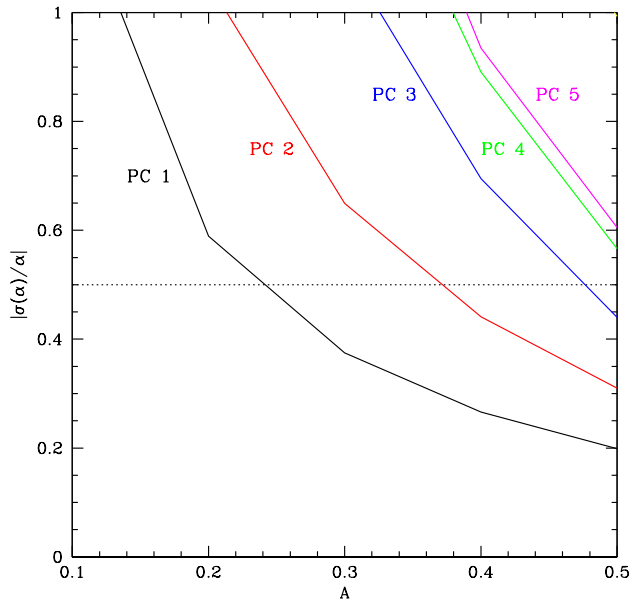,width=3.4in}
\caption{As Fig.~\ref{fig:sigalf}, but for $w(a)=-1+A\,[1-\cos(\ln a)]$ 
with non-monotonic behavior.  Note $w$ would exceed 0 at some redshifts 
for $A>0.5$. 
}
\label{fig:sigalf2}
\end{center}
\end{figure}

Figures~\ref{fig:sigalf}--\ref{fig:sigalf2} illustrate the physical 
discriminating power of this idealized experiment as a function of 
cosmological model.  For most models viable under current data, 
2-3 PCs have sufficient $S/N$ to be useful in distinguishing between 
the true cosmology and the cosmological constant.  Of course as the 
model approaches the cosmological constant, distinction becomes more 
difficult.

To exhibit robustness of these results against the 
specific assumed model, we scan over model parameters for classes of 
qualitatively different dark energy physics, representing the thawing 
class, freezing class, and a non-monotonic EOS.  In the last case, we 
see that the ability of PCA not to be locked into a particular 
functional form, e.g.\ a monotonic parametrization, does not make 
more degrees of freedom significant.  

Again, what we really care about is the $S/N$.  In Table~\ref{tab:sn2} 
we list the fraction of the total $S/N$ (Eq.~\ref{eq:snr}) 
contributed by the two best modes -- for the case 
for each dark energy class where higher modes contribute 
the {\it most\/}.  For the thawing class the higher modes add less 
than 0.3\% to the total, and for the freezing class less than 2.8\% in 
the most sensitive case, dropping to less than 0.5\% for modes above 
the third.  And recall this was for a highly idealized experiment.  In 
the oscillating model, an ad hoc case designed to be especially 
PCA-friendly, the most extreme case with oscillations reaching $w=0$ 
allows higher modes to contribute up to 14\% (8\% for above the third 
mode).  These are actually {\it overestimates\/} of the importance of high 
modes because as discussed in the next section the $S/N$ of the higher 
modes degrades when $w(z>9)$ is marginalized over rather than fixed.

\begin{table}[htbp]
\begin{center}
\begin{tabular*}{0.95\columnwidth} 
{@{\extracolsep{\fill}} l c c }
\hline
Model & $(S/N)_2/(S/N)_{\rm all}$ & \ $(S/N)_3/(S/N)_{\rm all}$ \\ 
\hline
Freezing ($w_a=0.7$)& 0.972& \ 0.995 \\ 
Thawing ($w_a=-0.5$) & 0.997& \ 0.9998 \\ 
Oscillating ($A=0.5$) & 0.862& \ 0.922 \\ 
\hline 
\end{tabular*}
\caption{Fraction of total signal-to-noise contributed by the first 
two, or three, principal components for the case in each dark energy 
class {\it most\/} favoring PCA high modes.} 
\label{tab:sn2}
\end{center}
\end{table}

Interestingly, fitting $w_0$-$w_a$ 
rather than using PCA provides distinction from the cosmological 
constant at the $1\sigma$ ($2\sigma$) level for $|w_a|=0.13$ (0.24) -- 
very comparable to the PCA approach.  That is, the second parameter 
becomes useful at almost the same values as in the top panel of 
Fig.~\ref{fig:sigalf}, for the PCA freezing case, and is more 
sensitive than in the bottom panel for the PCA thawing case. 
Both methods demonstrate that significant physical constraints on 
the dark energy EOS are described by of order two quantities. 

The main point though is that the important information is not in 
$\sigma_i$, but $\sigma_i/\alpha_i$.  Just because PCA may say 
uncertainties $\sigma_i$ are small, 
this does not mean that we know the physics answer.

\section{What Happens at High Redshifts, Stays at High Redshifts?} 

While the previous demonstration of PC uncertainties and 
signal-to-noise is 
the most important of this paper, we also note that assumptions about 
the high redshift EOS behavior have significant effects.  
In practice, the PCs are often computed assuming a cut off at some 
maximum redshift to avoid complications in calculating 
the cosmic microwave background (CMB) primordial power spectra and the 
initial conditions for growth of matter perturbations.  At higher 
redshifts one must therefore choose a particular form or value for 
the EOS.  FOMSWG fixes $w(z>9)=-1$.  One justification 
is that for the cosmological constant the dark energy density 
fades away quickly into the past, so the exact value of $w$ there is 
unimportant.  However, we have no guarantee that the cosmological 
constant is the correct model, nor even essentially any current 
information on the behavior of the EOS at $z>1$ \cite{kowalski}. 
Assumptions about the high redshift behavior can lead to significant 
biases and improper conclusions about the nature 
of dark energy (see, e.g., \cite{deplpca}).  

Bias should not be as severe a problem for a high transition redshift, 
$z=9$, and with many redshift bins at $z\gtrsim3$ the extra degrees of 
freedom should ameliorate bias from 
the prior at $z>9$.  However, fixing $w(z>9)=-1$ does demonstrably 
influence the PCs: for example some of the uncertainties 
$\sigma_i/\alpha_i$ that are apparently tightly determined 
can degrade by a factor three when $w(z>9)$ is not fixed. 
The modes themselves also change shape, as we discuss next. 

Thus, what happens at high redshift 
does {\it not\/} stay at high redshift, but can affect some important 
aspects of the principal component analysis.

\section{Highest is Best?} 

Does the location in redshift of the maximum of a PC, say the first one, 
say something fundamental about the science reach of the survey or probe 
employed?  No -- as is clear from Eq.~(\ref{eq:pc}) the EOS constraints 
follow from the sum -- with both positive and negative contributions -- 
over all the PCs, not any single one.  

\begin{figure}[!htb]
\begin{center}
\psfig{file=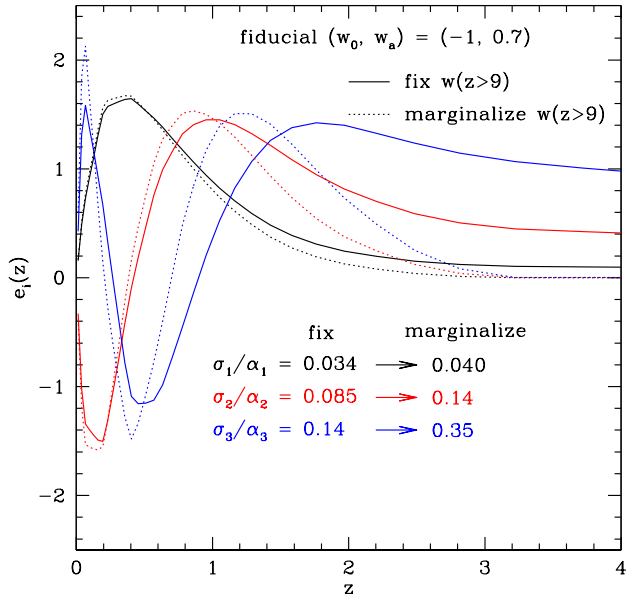,width=3.4in}
\caption{The location and shape of peaks in principal component modes 
depend on the high redshift treatment of the EOS.  The peak location 
therefore does not in itself translate into the science impact of a given 
experiment.  Note the effects are more severe for a more realistic 
experiment. 
}
\label{fig:pchiz}
\end{center}
\end{figure}

Moreover, Figure~\ref{fig:pchiz} demonstrates that artificially fixing 
the high redshift EOS behavior changes the PC shape and peak location.  
This shift has nothing to do with the experimental design and so the 
peak location is not a signpost to experiment optimization.  
Assumptions on $w(z>9)$ can affect probes differently: e.g.\ supernova 
distances do not involve $w(z>9)$ while baryon acoustic oscillations, 
being tied to high redshift, do.

Finally, even if a PC peak did mean that the experiment is most sensitive 
to the dark energy EOS at a higher redshift, say, that would not imply 
that the experiment is most sensitive to the nature of dark energy.  
For example, dark energy is most influential today, so perhaps one wants 
an experiment most sensitive to the low redshift behavior.   
(We emphasize that understanding dark energy at low redshift still 
requires measuring expansion and growth to high redshift, to break 
degeneracies.)  At best, one could say that probes that weight the dark 
energy differently in redshift have some complementarity.  

But there is 
no justification for claiming that the probe with the highest peak, or 
with the peak at the highest redshift, is the best probe.

\section{Conclusions} 

Principal component analysis is a valid technique, 
used appropriately.  Oversimplifying PCA interpretation or 
inadequately appreciating the effect of assumptions 
employed can lead to misunderstandings and false beliefs.  We 
present cautionary examples of three apparently plausible but 
unjustified extrapolations.  While data 
should be analyzed in every reasonable manner, for understanding the 
generic cosmology reach the more complicated PCA approach 
demonstrates no extraordinary advantage over the well-tested and highly 
calibrated phase space dynamics approach of $w_0$-$w_a$.

\acknowledgments 

We thank Andy Albrecht and Dragan Huterer for detailed discussions of 
PCA issues, and Bob Cahn for useful suggestions.  
This work has been supported in part by the Director, Office of Science, 
Office of High Energy Physics, of the U.S.\ Department of Energy under 
Contract No.\ DE-AC02-05CH11231.


\begin{thebibliography}{99}

\bibitem{fomswg} 
Figure of Merit Science Working Group summary, 
http://jdem.gsfc.nasa.gov/docs/Kolb\_JDEM\_FoMSWG.ppt 

\bibitem{detf} 
Dark Energy Task Force report, arXiv:astro-ph/0609591

\bibitem{calib} 
R. de Putter \& E.V. Linder, JCAP 0810, 042 (2008)

\bibitem{hutstar} 
D. Huterer \& G. Starkman, Phys. Rev. Lett. 90, 031301 (2003)

\bibitem{albern} 
A. Albrecht \& G. Bernstein, Phys. Rev. D 75, 103003 (2007)

\bibitem{deplpca} 
R. de Putter \& E.V. Linder, Astropart. Phys. 29, 424 (2008) 
 
\bibitem{mort} 
M.J. Mortonson, W. Hu, D. Huterer, arXiv:0810.1744 

\bibitem{linhut} 
E.V. Linder \& D. Huterer, Phys. Rev. D 72, 043509 (2005)

\bibitem{leach} 
S. Leach, MNRAS 372, 646 (2006) 

\bibitem{kadota} 
K. Kadota, S. Dodelson, W. Hu, E.D. Stewart, Phys. Rev. D 72, 023510 
(2005) 

\bibitem{kowalski} 
M. Kowalski et al., ApJ 686, 749 (2008) 

\end{thebibliography}
\end{document}